\documentclass[aps,prd,twocolumn,showpacs,amsmath,amssymb]{revtex4}

\usepackage{graphicx}
\usepackage{dcolumn}
\usepackage{bm}

\begin{document}

\title{Stability of nonlinear magnetic black holes}
\author{Nora Bret\'on}
\email{nora@fis.cinvestav.mx}
\affiliation{Departamento de F\'{\i}sica,
Cinvestav--IPN, Apartado Postal 14--740, C.P. 07000, M\'exico,
D.F., MEXICO}

\date{\today}  

\begin{abstract}
We study the stability of static spherically symmetric exact solutions
of Einstein equations coupled with nonlinear electrodynamics, in the
magnetic sector. These solutions satisfy the heuristic model proposed by
Ashtekar-Corichi-Sudarsky for hairy black holes, meaning that the
horizon mass is related to their Arnowitt-Deser-Misner (ADM) mass and to
the corresponding particle-like solution. We test the unstability
conjecture that emerges for hairy black holes and it turned out that it
becomes confirmed except for the Einstein-Born-Infeld solutions.
\end{abstract} 

\pacs{04.70.-s, 04.70.Bw, 04.70.Dy}
\maketitle


\section{Introduction}

Remarkable properties of nonlinear electrodynamical black holes arise in
the context of the isolated horizon formalism, recently put forward by
Ashtekar {\it et el} \cite{Ashtekar}.  In this approach it is pointed out
the unsatisfactory (uncomplete) description of a black hole given by
concepts such as ADM mass and event horizon, for instance, specially if
one is dealing with hairy black holes. To remedy this uncompleteness,
Ashtekar {\it et al} have proposed alternatively the isolated horizon
formalism, that furnishes a more complete description of what happens in
the neighborhood of the horizon of a hairy black hole.

In the isolated horizon formalism one considers spacetimes with an
interior boundary, which satisfy quasi-local boundary conditions that
insure that the horizon remains isolated. The boundary conditions imply
that quasi-local charges can be defined at the horizon, which remain
constant in time. In particular one can define a horizon mass, a horizon
electric charge and a horizon magnetic charge.

Moreover, Ashtekar-Corichi-Sudarsky (ACS) conjecture about the
relationship between the colored black holes and their solitonic analogs
\cite{Sudarsky}: the Arnowitt-Deser-Misner (ADM) mass contains two
contributions, one attributed to the black hole horizon and the other to
the outside hair, captured by the solitonic residue.  In this model, the
hairy black hole can be regarded as a bound state of an ordinary black
hole and a soliton.
 
For colored black holes the difference between the horizon and ADM masses
can be seen as the energy that is available for radiation to fall both
into the black hole and to infinity. In static solutions there is no
radiation, thus a positive value of the difference $M^{ADM}- M_{\Delta}=E$
means that a slightly perturbation on the initial data of the static
solution will lead the system to a Schwarzschild black hole and the total
radiated energy will be equal to $E$. For colored black holes a positive
$E$ is a necessary condition for the solution to be unstable.
   
It is then natural to ask if those conjectures do apply to the case of
(non colored) black hole solutions of Lagrangians more general than the
Einstein-Maxwell one.  In the present paper we address static spherically
symmetric (SSS) solutions of nonlinear electrodynamics (NED) coupled with
gravity, the corresponding action given by

\begin{equation}
S = \int{d^4x \sqrt{-g} \{ R (16\pi)^{-1}-L \} } ,
\end{equation}
where $R$ denotes the scalar curvature, $g:= {\rm det} \vert g_{\mu \nu}
\vert$ and $L$, the electromagnetic part, is assumed to depend in
nonlinear way on the invariants of $P_{\mu \nu}$, the nonlinear
generalization of the electromagnetic field, $F_{\mu \nu}=
\partial_{\mu}A_{\nu}-\partial_{\nu}A_{\mu}$,

\begin{equation}
L= \frac{1}{2} P^{\mu \nu} F_{\mu \nu} -K(F, \tilde{F}),
\label{Lagr}
\end{equation}
where $F$ and $\tilde{F}$ are the invariants of $P_{\mu \nu}$ and
$K(F, \tilde{F})$  is the structural function.

We verify the mass formula for static spherically symmetric black hole
solutions and their soliton-like counterpart for three distinct
nonlinear electrodynamics. We show that the masses relation is fulfilled
and that the value of the ``total energy" $E$ for SSS-NED solutions is
positive.

Moreover we test the stability of the solitons and black holes with
respect to nonspherically symmetric dynamical linear fluctuations using
the results obtained by Moreno and Sarbach. In \cite{M-S} the dynamical
stability of solutions in self-gravitating nonlinear electrodynamics is
analyzed with respect to arbitrary linear fluctuations of the metric and
electromagnetic field. Conditions are derived on the electromagnetic
Lagrangian which imply linear stability. We show that these conditions
hold for Einstein-Born-Infeld purely magnetic solutions but they fail
for other solutions of NED.
   
The Born-Infeld example leads to the conclusion that for SSS-NED black
holes and particle-like solutions the unstability conjecture derived
from the positive difference between the horizon and ADM masses does not
hold generically.

We shall restrict our attention to purely magnetic solutions because of
two reasons: firstly, there exists no go theorems \cite{Bronn} that
forbid the existence of purely electric or dyonic (electric and magnetic
charge) solutions with regular center. However the prohibition does not
concern purely magnetic solutions and there is a whole class of regular
solutions with a nonzero magnetic charge. Since we want to test the ACS
heuristic model we focus on NED theories possessing both solutions black
hole like and its solitonic counterpart.
 
The second reason to consider purely magnetic solutions is that for
purely magnetic cases the horizon mass can be calculated in a
straighforward way once the metric is known because the magnetic charge
does not contribute as a global charge;  the horizon mass is given by:
 
\begin{equation}
M_{\Delta} = \frac{1}{2}
\int_{0}^{r_{\Delta}}{[1-2M'(\tilde{r})]d\tilde{r}} ,
\end{equation}
where $M(r)$ is given by the metric function $N(r)=1-{2M(r)}/{r}$
in the SSS line element, 

\begin{equation}
ds^2=-N(r)dt^2+N(r)^{-1}dr^2+r^2(d
\theta^2 + \sin^2{\theta}d\phi^2),
\label{sss-metric}
\end{equation}

In contrast, for the electric case the expression for the horizon mass
depends on the theory:  it involves a potential,
$V(a_{\Delta},P_{\Delta},Q_{\Delta})$, that depends on the horizon
parameters and must be determined in consistence with the first law of
black hole mechanics \cite{ulises}:

\begin{equation}
M_{\Delta} = \frac{1}{4 \pi} \kappa a_{\Delta} + \Psi
Q_{\Delta}+V(a_{\Delta},P_{\Delta},Q_{\Delta}),
\end{equation}
where $\kappa$ is the surface gravity of the horizon, $a_{\Delta}$ is
the surface area, $\Psi$ is the electric potential, $Q_{\Delta}$ and
$P_{\Delta}$ are the electric and magnetic charges of the horizon,
respectively.

On the other side, in NED there are cases in which the known solution is
purely electric, however, for spherically symmetric solutions in
nonlinear electrodynamics there exist a duality that connects purely
electric solutions with purely magnetic ones.  The electromagnetic field
tensor $F_{\mu \nu}$ compatible with spherical symmetry can have a
radial electric field ($F_{01}$) and a radial magnetic field ($P_{23}$).
The FP duality let us to pass from a solution with a radial electric
field to a solution with a radial magnetic field and viceversa; the two
solutions corresponding to the same line element (same $g_{\mu \nu}$)
but to different Lagrangians. The FP duality is described in more detail
in \cite{Bronn}.

We shall address three solutions magnetically charged in nonlinear
electrodynamics coupled to gravity, all of them having both types of
solutions, particle-like and black hole ones. The studied cases are: 1)
The purely magnetic Born-Infeld black hole and its corresponding EBI-on.
2)The magnetic solution obtained recently by Bronnikov \cite{Bronn}, via
FP duality, from a purely electric one derived by Ayon-Garcia
\cite{Ayon}. 3) A magnetic solution determined via FP duality from a
regular electric solution recently presented in \cite{Dymni}; this last
solution avoided the non existence theorems by having a de Sitter
center. Each case corresponds to a different nonlinear electrodynamical
Lagrangian.

\section{Born-Infeld black hole and EBIon}

The EBI black hole is the solution for the field equations of the
Lagrangian in (\ref{Lagr}) with the Born-Infeld structural function
given by

\begin{equation}
K=b^2 \left(1-\sqrt{1-{2F}/{b^2}+ {{\tilde{F}}^2}/{b^4}} \right),
\label{BIK}
\end{equation}
where $b$ is the maximum field strenght and the relevant parameter of
the BI theory. For the purely magnetic case one of the two
electromagnetic invariants constructed from $F_{\mu \nu}$ is zero
(${\tilde{F}}=0$).
  
The EBI solution for a SSS spacetime as (\ref{sss-metric}) is given
by the metric function

\begin{eqnarray}
N(r)&=& 1-\frac{2m}{r} + \frac{2}{3}b^2 (r^2 - \sqrt{r^4+a^4})+
\frac{4g^2}{3r}G(r), \\
G'(r)&=&- (r^4+a^4)^{- \frac{1}{2}},
\label{BImetrfunc}
\end{eqnarray}
where $G'(r)$ denotes the derivative of $G(r)$ with respect to the
radial variable, $m$ is the mass parameter, $g$ is the magnetic charge
(both in lenght units), $a^4=g^2/b^2$ and $b$ is the Born-Infeld
parameter given in units of $[\rm{lenght}]^{-1}$. The nonvanishing
components of the electromagnetic field are

\begin{equation} 
F_{rt}= g (r^4+ a^4)^{- \frac{1}{2}},
\quad P_{rt}= \frac{g}{r^2}. 
\label{FrtBI}
\end{equation} 
 
The black hole solution given by Gar\-c\'{\i}a-Sa\-la\-zar-Ple\-ba\~nski
\cite{GSP} corresponds to

\begin{equation}
G(r)= \int^{\infty}_{r}{\frac{ds}{\sqrt{s^4+a^4}}}=\frac{1}{2a}
{\mathbb{F}} \left[ \arccos{ \left( \frac{r^2- a^2}{r^2+a^2} \right)},
\frac{1}{\sqrt{2}} \right],
\label{gPleb}
\end{equation}
where ${\mathbb{F}}$ is the elliptic integral of the first kind. On
the other side, the particle-like solution given by Demianski
\cite{Geon} is

\begin{equation} 
G(r)=  \int^{r}_{0}{\frac{-ds}{\sqrt{s^4+a^4}}}=- \frac{1}{2a}
{\mathbb{F}} \left[ \arccos{ \left(
\frac{a^2-r^2}{a^2+r^2}\right)},\frac{1}{\sqrt{2}} \right].
\label{gDem}
\end{equation}
     
The election of $G(r)$ as in Eq. (\ref{gPleb}) or Eq. (\ref{gDem}) has
as a consequence a different behavior of the solution at the origin.  
The metric function $N(r)$ with $G(r)$ given by Eq. (\ref{gPleb})  
diverges at $r \to 0$ (even when $m=0$), corresponding to the black hole
solution. The other one, meaning $N(r)$ with $G(r)$ given by Eq.
(\ref{gDem}) is the so called EBIon solution that is finite at the
origin (for $m=0$). The integrals of Eqs. (\ref{gPleb}) and (\ref{gDem})
are related by

\begin{equation}
\int^{\infty}_{r}{\frac{ds}{\sqrt{s^4+a^4}}}+\int^{r}_{0}
{\frac{ds}{\sqrt{s^4+a^4}}}=\frac{1}{a}{\rm K} \big[\frac{1}{2} \big],
\end{equation}
where ${\rm K}[\frac{1}{2}]$ is the complete elliptic integral of the
first kind. In the limit of large distances, $r \to \infty$,
asymptotically the solution corresponds to the magnetic
Reissner-Nordstrom (RN) solution, the SSS solution to Einstein-Maxwell
equations. Also when the BI parameter goes to infinity, $b \to
\infty$, we recover the linear electromagnetic (Einstein-Maxwell) RN
solution. In the uncharged limit, $b=0$ (or $g=0$), it is recovered
the Schwarzschild black hole.

\subsection{The mass relation}

When the black hole is not completely determined by global charges
defined at spatial infinity such as ADM mass, angular momentum or
electric charge, but rather it possesses short range charges (hair) that
vanish at infinity, then it is a hairy black hole. In a series of papers
Ashtekar {\it et al} have proposed a more complete description to
characterize a hairy black hole, based on quantities defined at the
horizon. This formalism is intended to deal with situations more general
than SSS hairy black holes and it involves the canonical formalism of
gravity.  Furthermore, they proposed a formula relating the horizon mass
and the ADM mass of the colored black hole solution with the ADM mass of
the soliton solution of the corresponding theory,

\begin{equation}
M^{(n)}_{sol}=M^{(n)}_{ADM} -M^{(n)}_{\Delta},
\label{massrel}
\end{equation}
 
\noindent where the superscript $n$ indicates the colored version of the
hole; in the papers of Ashtekar {\it et al} this $n$ refers to the
Yang-Mills hair, labeled by this parameter, corresponding to $n=0$ the
Schwarszchild limit (absence of YM charge). This relation has been
proved numerically to work for the Einstein-Yang-Mills (EYM) black hole.
 
It was shown in \cite{Breton} that the EBI black hole and the
corresponding EBIon solution fulfill the relation between the masses as
well as most of the properties of the model for the colored black hole.
For the EBI solution the horizon and ADM masses as functions of the
horizon radius $r_{\Delta}$ are given, respectively, by

\begin{widetext}
\begin{equation} 
M^{(b)}_{\Delta}(r_{\Delta})= \frac{r_{\Delta}}{2}+\frac{b^2
r_{\Delta}}{3}(r_{\Delta}^2-\sqrt{r_{\Delta}^4+a^4})-\frac{2g^2}{3}
\int^{r_{\Delta}}_{0}{\frac{ds}{\sqrt{a^4+s^4}}}, 
\label{Mhor}
\end{equation}

\begin{equation} 
M^{(b)}_{ADM}(r_{\Delta})= \frac{r_{\Delta}}{2}+\frac{b^2
r_{\Delta}}{3}(r_{{\Delta}}^2-\sqrt{r_{{\Delta}}^4+a^4})+\frac{2g^2}{3}
\int^{\infty}_{r_{\Delta}}{\frac{ds}{\sqrt{a^4+s^4}}},
\label{Madm}
\end{equation}
\end{widetext}

The mass of the soliton can be obtained by letting $r_{\Delta} \to 0$
in the ADM mass, Eq. (\ref{Madm}), obtaining $M_{sol}^{(b)}=2g
\sqrt{gb}{\rm K}[\frac{1}{2}]/3$. From these expressions one can
trivially check that they satisfy Eq. (\ref{massrel}).
 
\subsection{Stability of the EBI solutions}
 
For hairy black holes the positivity of the difference between the ADM
mass and the horizon mass, $M_{ADM}-M_{\Delta}=E >0$, indicates that
there exists an energy $E$ available to be radiated. For static black
holes this result can be interpreted as a potential unstability, i.e. a
slightly perturbation in the initial data will lead the solution to
decay to a Schwarzschild black hole.

Stability properties in self-gravitating nonlinear electrodynamics were
studied by Moreno and Sarbach \cite{M-S}.  They derived sufficient
conditions for linear stability with respect to arbitrary linear
fluctuations in the metric and in the gauge potential, $\delta g_{\mu
\nu}$ and $\delta A_{\mu \nu}$, respectively; the conditions were
obtained in the form of inequalities to be fulfilled by the nonlinear
electromagnetic Lagrangian $L(F)$ and its derivatives.

The application of this criterion is restricted to static, spherically
symmetric solutions of NED coupled to gravity, that are purely electric
or purely magnetic. For these systems a gauge invariant perturbation
formalism was used obtaining that linear fluctuations around a SSS
purely electric (or purely magnetic) solution are governed by a wavelike
equation with symmetric potential of the form:

\begin{equation}
(-P \tilde{\nabla}^aP^{-1} \tilde{\nabla}_a P+S)u=0,
\end{equation}
where $\tilde{\nabla}$ denotes covariant derivative of the metric
$\tilde{g} =-Ndt^2+N^{-1}dr^2$, $P$ is a positive-definite symmetric
matrix and $S$ is a symmetric matrix; $u$, a vector-valued function, is
a gauge-invariant combination of the perturbed metric and perturbed
electromagnetic field. Linear stability follows if the potential $S$ is
positive-definite. In terms of the Lagrangian $L(F)$ and metric function
$N$, the positiveness of $S$ is accomplished if

\begin{equation}
L(y) >0, \quad L(y)_{,y} >0, \quad L(y)_{,yy} >0 .
\label{stabcond1}
\end{equation}
where $y= \sqrt{2g^2F}$ ($g$ is the magnetic charge and $F$ is one of
the invariants of the electromagnetic field.
  
Besides, there are more inequalities to be fulfilled, that arise from
the pulsation equations for $l \ge 2$ in the even-parity sector.  The
potential $S$ is positive definite if $\kappa > 0$, $N \le 1$ and also
$ l (l+1) - 2N \kappa >0$. While for $l=1$ the corresponding potential
is positive if $\kappa >0$, where $\kappa$ is defined as

\begin{equation}
\kappa=1+ 2 L_{,F}^{-1}L_{,FF}F,
\end{equation}
$L_{,F}$ denotes the derivative of the Lagrangian with respect
to the invariant $F$. In terms of the variable $y= \sqrt{2g^2F}$ these
additional inequalities are equivalent to

\begin{equation}
f(y)= yL_{,yy}/L_{,y}>0, \quad
f(y)N(y)< 3.
\label{stabcond2}
\end{equation}

We shall apply this criterion to test the stability of the particle-like
and black hole solutions; in the former case the boundary point is the
origin, $r=0$, while for the black hole case the conditions must be held
in the domain of outer communication (DOC), i.e. positions outside the
horizon, $r> r_h$, $r_h$ being the radius of the horizon of the black
hole.

The BI Lagrangian fulfill the stability conditions; in terms of the
variable $y$, the BI Lagrangian, Eq. (\ref{Lagr}), is given by

\begin{equation}
L(y)=b^2[\sqrt{1+\frac{y^2}{b^2g^2}}-1]>0,
\end{equation}
and the rest of the inequalities read as:

\begin{eqnarray}
L_{,y}&=&\frac{y}{g^2}(1+\frac{y^2}{b^2g^2})^{-\frac{1}{2}}>0,
\nonumber\\
L_{,yy}&=&\frac{1}{g^2}(1+\frac{y^2}{b^2g^2})^{-\frac{3}{2}}>0,\nonumber\\
f(y)&=& y\frac{L_{,yy}}{L_{,y}} =(1+\frac{y^2}{b^2g^2})^{-1}>0,
\label{BIineq}
\end{eqnarray}
 
Conditions (\ref{BIineq}) are fulfiled in all the range of $y$.
Moreover, $f(y)$ is monotonically decreasing with $f(y=0)=1$, $0<f(y)
\le 1$; then the last stability condition $f(y)N(y) < 3$ reduces to
prove that $N(y)< 3$, for the black hole it must be fulfilled in DOC,
while for the particle-like solution the domain to be considered is $0
\le r < \infty$.
      
In the black hole case, the metric function $N(r)$ has a minimum in the
extreme case ($g=m$) for $bm=0.5224$ at $r_h=0.346 m$; DOC is considered
for distances larger than the radius of the horizon, $r> r_h=0.346 m$.
In terms of $y$, considering that $F=g^2/2r^4$ then $y=g^2/r^2$, the
metric function $N(y)$ is

\begin{eqnarray}
N(y)&&=1-\frac{2m \sqrt{y}}{g}+
\frac{2b^2g^2}{3y}[1-\sqrt{1+\frac{y^2}{b^2g^2}}]+ \nonumber\\
&&\frac{2 \sqrt{gby}}{3}{\mathbb{F}} \left[
\arccos({\frac{gb-y}{gb+y}}),
\frac{1}{\sqrt{2}} \right],
\end{eqnarray}
  
In the range $0<y<y_h =8.35$ it turns out that $0<N(y) \le 1$ with
$N(0)=1$ therefore, $0<N(y)<1<3$, fulfilling the last inequality
required as sufficient conditions for stability of the EBI black hole.
  
For the particle-like solution of the EBI equations, the metric
function $N(r)$ in terms of $y$ is

\begin{eqnarray}
N(y)&&=1-\frac{2m \sqrt{y}}{g}+
\frac{2b^2g^2}{3y}[1-\sqrt{1+\frac{y^2}{b^2g^2}}]- \nonumber\\
&&\frac{2 \sqrt{gby}}{3}{\mathbb{F}} \left[
\arccos({\frac{y-gb}{gb+y}}),
\frac{1}{\sqrt{2}} \right],  
\end{eqnarray}
  
The last stability condition $N < 3$ in fact occurs since $N(y=0)=1$ and
the function is monotonically decreasing, having $N(y) \le 1$, for $bg
\ne 0$; the finiteness in the origin of $N(r)$ is valid when $m=0$.
Therefore, as far as this analysis proves, the EBI solutions, both black
hole and particle-like one, are stable, becoming in this case non valid
the unstability conjecture. However, this situation is far from being
generic in NED as the next two examples will show.

\section{Bronnikov solution}
 
In \cite{Bronn} were derived magnetic black hole and soliton-like
solutions with NED coupled to gravity. The Lagrangian in this case is
of the form

\begin{equation}
L=F/ \cosh^2{(a \vert{F/2}\vert ^{1/4})},
\label{Lbronn}
\end{equation}
where $a$ is a constant, $F=2g^2/r^4$, $g$ is the magnetic charge .
The use of $\vert{F}\vert$ violates analyticity of $L$ at $F=0$,
though, in the range of interest, $F>0$, $L(F)$ is well behaved. The
corresponding function $M(r)$ in the line element (\ref{sss-metric})
is given by

\begin{equation} 
M(r)=\frac{g^{\frac{3}{2}}}{2a}[1-\tanh(\frac{a \sqrt{g}}{r})],
\end{equation}

We also have, from the Komar integral, that

\begin{equation}
M_{ADM}(r)=   
\frac{g^{3/2}}{2a}[1-\tanh{(\frac{a \sqrt{g}}{r})}]-
\frac{g^{2}}{2r} {\rm sech}^2{(\frac{a \sqrt{g}}{r})},
\end{equation}

\begin{equation}
\lim_{r \to \infty} M_{ADM}=\frac{g^{3/2}}{2a},
\end{equation}  

Moreover, the minimum value of $N(r)=1-2M(r)/r$ depends on the ratio
$\xi=m/g$ (we consider $g>0$), so that $N_{min}$ is negative for $\xi >
\xi_{0} \approx 0.96$ (we deal with a black hole with two horizons),
zero for $\xi = \xi_{0}$ (an extremal black hole with one double
horizon) and positive for $\xi < \xi_{0}$ (a regular particle-like
system).Given any specific value of the constant $a$ in
Eq.(\ref{Lbronn}) we can obtain all three types of solutions depending
on the charge value; we have a nonextremal or extremal black hole if $g
\le 4a^2{\xi_o^2}$, or we have a particle-like solution (a monopole)
otherwise.

As function of the horizon radious $r_{\Delta}$ the ADM mass and the
horizon mass are, respectively,

\begin{eqnarray}
M_{ADM}(r_{\Delta})&=&\frac{r_{\Delta}}{2}-  
\frac{g^{3/2}}{2a}[1-\tanh(\frac{a \sqrt{g}}{r_{\Delta}})]
+\frac{g^{3/2}}{2a}, \nonumber\\
M_{\Delta}(r_{\Delta})&=&\frac{r_{\Delta}}{2}-
\frac{g^{3/2}}{2a}[1-\tanh(\frac{a \sqrt{g}}{r_{\Delta}})],
\label{mbronn}
\end{eqnarray}

The limits when the horizon radius go to zero, $r_{\Delta} \to 0$, are

\begin{equation}
\lim_{r_{\Delta} \to 0}M_{\Delta}(r_{\Delta})=0, \quad
\lim_{r_{\Delta} \to 0}M_{ADM}(r_{\Delta})=\frac{g^{3/2}}{2a},
\end{equation}
  
From expressions (\ref{mbronn}) it can be shown that the relation
(\ref{massrel}) holds:
$M_{ADM}(r_{\Delta})-M_{\Delta}(r_{\Delta})=g^{3/2}/2a$, obtaining the
soliton mass $g^{3/2}/2a$. According to the hairy black hole heuristic
model, the solution may be unstable since $M_{ADM}>M_{\Delta}$. This
turns out to be the case as will be shown below.

The sufficient conditions Eqs. (\ref{stabcond1}) and (\ref{stabcond2})
for the solution to be stable are fulfilled for the black hole in the
domain of outer communication, however the particle-like solution fails
to hold most of the stability conditions. In terms of the variable $y=
\sqrt{2g^2F}$, the Lagrangian (\ref{Lbronn})  is
 
\begin{eqnarray}
L(y)&=&\frac{y^2}{2g^2} {\rm sech} ^2{x}>0,\nonumber\\
L_{,y}&=&\frac{y}{g^2} {\rm sech}^2{x}[1-\frac{x}{2}
\tanh{x}]> 0, 
\nonumber\\ 
L_{,yy}&=& \frac{{\rm sech}^2{x}}{g^2} [
1-\frac{x^2}{4}-\frac{7}{4}x \tanh{x}+\frac{3x^2}{4}
\tanh^2{x}]>0,\nonumber\\
f(y)&=&y L_{,yy}/L_{,y}>0.
\label{ineq}
\end{eqnarray}

where $x=a \sqrt{y/2g}$. The condition $L_{,y}>0$ is fulfilled if
$2-x{\tanh{x}}>0$, or $x < 2.065$. That in terms of $y$ amounts to $y <
2(2.065)^2 g/a^2$. If we consider the black hole case, for which $g/a^2
< 4 {\xi_o^2}$, then this condition is fulfilled if $y < 2(2.065)^2 4
{\xi_o^2}= 31.439$. In terms of $r$, $L_{,y}>0$ is held if $r_h \ge r >
g/(2(2.065) {\xi_o})=(0.252)g$, with $r_h$ being the horizon radius. To
define if this range is contained in DOC, we analyze the metric function
$N(r)$; in terms of $y$ it is

\begin{equation}
N(y)=1-\sqrt{\frac{yg}{2a^2}}[1 -
\tanh{(a \sqrt{\frac{y}{2g}})}].
\end{equation}

Or in terms of the variable $x=a \sqrt{y/2g}$, $N(x)=1-x(1-
\tanh{x})g/a^2$; independently of the nonvanishing value of $a$ and $g$,
$N(x)$ has a single minimum (double horizon) for $x_h=0.64$ that is the
only positive solution for $x$ in $ \partial_x N(x)=0$. This value
amounts to $g/a^2=3.571$. The corresponding value is $y_h=2.93$, or
since $y=2g^2/r^2$, $r_h= 0.827 g$. Putting it together then, $r_h=0.827
g > r > 0.252 g$, so $L_{,y}>0$ in DOC.

In relation to the stability condition $L_{,yy}>0$, independently of the
values of the constants, it holds in the range $x < 0.886$ and $x >
2.827$. In other words, the condition fails to hold in the range $ 0.886
< x < 2.827 $. The inequality $x < 0.886$ is equivalent, in terms of
$r$, to $r>\frac{g}{0.886} \sqrt{\frac{a^2}{g}}$ that for the black hole
can be casted as $r>{g}/((0.886)2{\xi_o})=0.587g$; then for the black
hole $L_{,yy}>0$ since $r_h=0.827 g > r > 0.587 g$.
  
There is a fourth condition still to be fulfilled, $3 > N(y)f(y)$.  
Since $ 0 < \tanh{x} \le 1, \quad x > 0$ then $0<N(y)<1$ in DOC; also
$N(y) f(y) >0$. Moreover, in the range $y < y_h=2.93$, $N(y) f(y) \le
1$, satisfying the inequality $3> N(y) f(y) >0$;  holding then all the
sufficient conditions for the black hole stability.

However, for the particle-like solution it can be shown that the
stability condition $L_{,y}> 0$ fails to hold. If $y > 2(2.065)^2g/a^2$
then $L_{,y}< 0$, or in terms of $r$ if $r < g/(2(2.065){\xi_o})$, that
evidently can occur when $r$ approaches the origin. The condition for
$L_{,yy}< 0$, $ x > 0.886$ that amounts to $r < g/(1.772 {\xi_o})$ can
also be attained when $r$ approaches the origin.
  
Since two of the conditions for stability are not fulfilled, in
particular when $L_{,y}<0$ and $L_{,yy}<0$ the consequence is that the
potential in the pulsation equations becomes negative and then
perturbations can grow without limit, we conclude that the particle-like
solution is unstable. A slightly perturbation of the soliton will lead
the system to a Schwarzschild black hole and the total radiated energy
will be equal to the diference between the ADM mass of the black hole
minus its horizon mass. Therefore we have shown that in this case the
unstability conjecture is valid.


\section{Dymnikova solution}
 
In \cite{Dymni} it was presented a SSS solution of NED coupled to
gravity that satisfies the weak energy condition (WEC) and is an
electrically charged regular structure. Discarding the requirement of a
weak field limit at the center, with a de Sitter center, the no go
theorems for electrically charged regular structures can be avoided. In
NED coupled to general relativity each electric solution has its
magnetic counterpart. We shall address here the magnetic structure
associated with Dymnikova's electric solution, obtained via FP duality.
 
The Lagrangian for the magnetic case is of the form

\begin{equation}
L=F/ (1+ \alpha \sqrt{F})^{2}
\end{equation}
where $\alpha=r_0^2/g \sqrt{2}$, $r_0=\pi g^2/8m$, $F=2 g^2/r^4$ and
$g$ is the magnetic charge.  The Lagrangian has stress-energy tensor
with the algebraic structure $T^t_t=T^r_r$; WEC leads to de Sitter
asymptotic at approaching a regular center. The corresponding function
$M(r)$ in the SSS line element is

\begin{equation} 
M(r)=\frac{2m}{\pi}[ \arctan(\frac{r}{r_0})-\frac{r r_0}{r^2+r_0^2}],
\end{equation}

As $r \to \infty$ it behaves as a Reissner-Nordstrom (RN) solution; for
$r << r_0$ it is asymptotically de Sitter with a cosmological constant
$\Lambda =g^2/2r_0^4$, $N(r \to 0)=1-\Lambda r^2/3$. The magnetic field
is given by
 
\begin{equation} 
B^2=\frac{g^2r^8}{(r^2+r_0^2)^6},
\end{equation}
 
In terms of $g/2m$ black hole exists for $g/2m \le 0.536$ while if $g/2m
> 0.536$ we have an electrically charged self-gravitating particle-like
NED structure. The metric function $N(r)$ has a double zero (double
horizon) for $r_{\pm}=1.825 r_0$. We also have, from the Komar integral,
the ADM mass and the soliton mass $m$ given by

\begin{equation}
M_{ADM}(r)=\frac{2m}{\pi}[ \arctan(\frac{r}{r_0})-
\frac{rr_0}{(r^2+r_0^2)}]-
\frac{q^2r^2}{2(r^2+r_0^2)^2},
\end{equation}

\begin{equation}
\lim_{r  \to \infty}M_{ADM}=m,
\end{equation}

For the magnetic version of Dymnikova's solution, the ADM mass
and the horizon mass, as functions of the horizon radius
$r_{\Delta}$, are, respectively,

\begin{eqnarray} 
M_{ADM}(r_{\Delta})&=&\frac{r_{\Delta}}{2}-
\frac{2m}{\pi}[ \arctan(\frac{r_{\Delta}}{r_0})-\frac{r
r_0}{r^2+r_0^2}]+m, \nonumber\\
M_{\Delta}(r_{\Delta})&=&\frac{r_{\Delta}}{2}-
\frac{2m}{\pi}[ \arctan(\frac{r_{\Delta}}{r_0})-\frac{r 
r_0}{r^2+r_0^2}],
\label{Dmasses}
\end{eqnarray}

The limits when the horizon radious goes to zero are

\begin{equation} 
\lim_{r_{\Delta} \to 0}M_{\Delta}(r_{\Delta})=0, \quad  
\lim_{r_{\Delta} \to 0}M_{ADM}(r_{\Delta})=m,
\end{equation}

From the previous expressions (\ref{Dmasses}) the difference between the
ADM mass and the horizon mass is positive, corresponding to the
particle-like mass $m$, $M_{ADM}(r_{\Delta})-M_{\Delta}=m$. Satisfying
in this way the ACS mass relation conjecture; the question arises if it
is related with black hole unstability. In terms of the variable $y=
\sqrt{2g^2F}$, according to the analysis in \cite{M-S} the conditions on
the electromagnetic Lagrangian which imply linear stability are

\begin{eqnarray}
L(y)&&=\frac{y^2}{(g \sqrt{2}+\alpha y)^{2}}>0,\nonumber\\
L(y)_{,y}&&=\frac{2 \sqrt{2}gy}{(g \sqrt{2}+\alpha
y)^{3}}>0,\nonumber\\
L(y)_{,yy}&&=\frac{4g(g- \sqrt{2} \alpha y)}{(g \sqrt{2}+\alpha
y)^{4}}>0,\nonumber\\
f(y)&&=y L_{,yy}/L_{,y}=\frac{\sqrt{2}(g- \sqrt{2}\alpha y)}{\sqrt{2}g
+ \alpha y}>0.
\label{Dymineq}
\end{eqnarray}
 
The last two inequalities hold if $y<g/ \sqrt{2} \alpha$. It is the case
for the black hole solution that $y<g/ \sqrt{2} \alpha$ in DOC.  The
double horizon occurs when $N(r)=0$, with two possible roots,
$r_{\pm}=1.825 r_0$; DOC involves distances $r>r_{+}$, or in terms of
$y$,

\begin{equation} 
y<\frac{2g^2}{r_{+}^2}=
\frac{2g^2}{(1.825 r_0)^2}= \frac{0.6 g}{\sqrt{2} \alpha},
\end{equation}

Since $0.6g/\sqrt{2} \alpha <g/ \sqrt{2} \alpha$ all the inequalities in
(\ref{Dymineq}) are held. To complete the stability test for the black
hole stability it must be shown that $3 > N(y) f(y)$.  The metric
function $N(y)$ is
 
\begin{equation} 
N(y)=1-\frac{g}{2r_0} \sqrt{\frac{y}{2}} \arctan \left( \frac{g}{r_0}
\sqrt{\frac{2}{y}} \right) +\frac{y}{2}
[2+y(\frac{r_0}{g})^2]^{-1}.
\end{equation}

If one considers that the extreme black hole occurs if $g/
\sqrt{\alpha}= g^2 \sqrt{2}/r_0^2= \sqrt{2} (2.37)^2$, then one obtains
$y_h=3.37$. In the interval $0<y <y_h=3.37$, the range of $N(y)$ is
$0<N(y)<1$. Moreover, since $N(0)=1$ and $f(0)=1$, the following
inequality holds $0<N(y)f(y)<1 < 3$. In such a manner that this magnetic
black hole has passed the stability test.

However, for the particle-like solution a simple analysis shows that
neither $L{,yy} >0$ nor $f(y) >0$ since $y < g/ \sqrt{2} \alpha$ amounts
to $r > \pi g (0.536)/2^{\frac{3}{2}}$ that fails to hold ($g \ne 0$)  
when $r$ approaches the origin ($r \to 0$).  We can check that the total
ADM mass of the black hole in Eq. (\ref{Dmasses}) is less than the sum
of the ADM mass of the soliton ($m$) and the horizon mass of the
Schwarzschild black hole ($r_{\Delta}/2$). If the NED soliton is
unstable, then, if perturbed, its mass can be radiated away and one ends
up with a bare black hole. In this case the NED black hole is considered
as unstable, confirming the ACS unstability conjecture.

\section{Final Remarks}
    
We have shown for three magnetic NED (coupled to gravity)  SSS solutions
that the ADM mass is greater than the horizon mass, $M_{ADM}^{(b)} >
M_{hor}^{(b)}$.  Since the diference between the hamiltonian horizon
mass and the ADM mass can be seen as the energy that is available for
radiation to fall both into the black hole and to infinity, then the
nonzero value of the Hamiltonian could be an indication of unstability
of the NED studied solutions. On this basis, one can conjecture that NED
black holes are unstable.  We have shown that this conjecture is valid
for some NED solutions, failing to be true in the case of the
Einstein-Born-Infeld black hole and its corresponding particle-like
solution, since the EBI solutions fulfill the sufficient stability
conditions concluding then that the unstability conjecture does not
apply generically for NED solutions.
  
We must mention that in relation to Born-Infeld black hole stability
recently was presented in \cite{Fernando} the analysis of quasinormal
modes for the gravitational perturbations, deriving a one dimensional
Schrodinger type wave equation for the axial perturbations. From the
behavior of the potentials it was concluded in \cite{Fernando} that the
EBI black holes are classically stable.
   
The stability problem deserves a deeper analysis in relation to the
energy conditions satisfied by the energy-momentum tensor since for the
NED unstable solutions their corresponding energy-momentum tensor
fulfills the weak energy condition only, while in the stable (EBI) case,
the dominant energy condition is satisfied by the EBI energy-momentum
tensor. Another feature worth to be mentioned is that the EBI black hole
is a singular one, while the unstable cases (Bronnikov and Dymnikova
solutions) correspond to regular black holes and the behavior of the
solution as black hole or as particle-like depends on the value given to
one parameter.


\end{document}